\documentclass[%
 preprint,
superscriptaddress,
showkeys,
 amsmath,amssymb,
 aps,
]{revtex4-1}

\usepackage{graphicx}
\usepackage{dcolumn}
\usepackage{bm}

\begin{document}

\preprint{APS/123-QED}

\title{Dynamics of Hydrogen Bonds Coupling on the Specific DNA-Protein Interactions}

\author{D. \surname{Dwiputra}}
 \email{donny.dwiputra@s.itb.ac.id}
 \affiliation{Theoretical Physics Laboratory, Faculty of Mathematics and Natural Sciences, Institut Teknologi Bandung, Jl. Ganesha 10, Bandung 40132, Indonesia}
\author{W. \surname{Hidayat}}%
 \email{wahid@fi.itb.ac.id}
 \affiliation{Theoretical Physics Laboratory, Faculty of Mathematics and Natural Sciences, Institut Teknologi Bandung, Jl. Ganesha 10, Bandung 40132, Indonesia}
 \affiliation{Indonesian Center of Theoretical and Mathematical Physics (ICTMP), Indonesia}
\author{F. P. \surname{Zen}}
\email{fpzen@fi.itb.ac.id}
 \affiliation{Theoretical Physics Laboratory, Faculty of Mathematics and Natural Sciences, Institut Teknologi Bandung, Jl. Ganesha 10, Bandung 40132, Indonesia}
 \affiliation{Indonesian Center of Theoretical and Mathematical Physics (ICTMP), Indonesia}

\date{\today}

\begin{abstract}
We propose a dynamical model depicting the interactions between DNA and a specific binding protein involving long range transmissions. The dynamics rely on the coupling between Hydrogen bonds formed between DNA and protein and between the base pairs because they account for site specificity of the binding. We adopt the Morse potential with coupling terms to construct the Hamiltonian. This model gives rise to a breather excitation, corresponding to the DNA bubble formation, which propagates as the carrier of genetic information. We examine the various kind of possible coupling dynamics and suggest the model feasibility in depicting the renaturation or hybridization processes.
\begin{description}
\item[PACS numbers]
87.15.-v, 05.45.-a, 52.35.Sb
\end{description}
\end{abstract}

\maketitle


\section{Introduction}
Life is fundamentally constructed by deoxyribonucleic acid (DNA), proteins, and their complex molecular interactions such as in gene regulation, transcription, and replication processes. These processes are essentially conducted by proteins that bind with very specific DNA sequences and communicate each other very efficiently even when separated by vast distances. Many models of the have been proposed \cite{ptashne1985gene,wang1988action,hogan1979transmission} to explain the nonlocal action at a distance. In this paper, our aim is to investigate the dynamics of a propagating local conformational distortion in DNA that acts as the information carrier triggered by the binding of protein in specific DNA sites. The dynamics of DNA recognition process by the protein is also discussed.

The study of nonlinear localized solitonic wave that propagates through DNA is initially conducted by Englander \textit{et al.} \cite{englander1980nature}, later by Peyrard and Bishop (PB) \cite{peyrard1989statistical} with their notable breather excitation refered as DNA bubbles, and by Yakushevich \cite{yakushevich1989nonlinear}. The dynamics and thermal effects of PB breather are studied in \cite{sulaiman2012dynamics,sulaiman2012thermal} while the effects of viscosity and external forces is also investigated \cite{hidayat2015viscous}. The PB model has been modified to include the protein interaction, such as the statistical model given in \cite{sataric2002impact} and TFAM-DNA interaction model \cite{traverso2015allostery} that depicts the allosteric interactions by DNA bubbles. However, the effect of the chemical bonds between protein and DNA to an analytical breather is not yet investigated \cite{donnyunpublishednonlinear}. The modelling of these chemical bonds is crucial as the specificity of the interaction are heavily dependent of these.

Origin of the specificity in the DNA-protein recognition consists of the complex chemical signatures carried by the base pairs and the sequence-independent DNA shapes \cite{rohs2010origins,halford2004site,von1986specificity}. Proteins "feel" the DNA surface, while simultaneously driven, by the electrostatic, van der Waals', and hydrogen bond (H-bond) interactions. The H-bonds role the specificity more significantly than the other interactions because the transition free energy between the most specific binding and the nonspecific case is approximately 16 $k_\mathrm{B} T$ below the specific binding energy \cite{williams2010biophysics}. For instance, the gap is experimentally 17 $k_\mathrm BT$ for \textit{Mnt} and $\approx$16 $k_\mathrm BT$ for \textit{lac} repressor \cite{gerland2002physical}. While, as a comparison, the value for nonspecific free energy is only about 7 $k_\mathrm BT$ for CI repressor protein in $\lambda$ virus-infected \textit{E. coli} cells in vivo \cite{williams2010biophysics}. 

We recall that every protein consists of amino acids and peptide bonds, the part of protein that interact directly with a DNA base is the side chain. When protein is near DNA, a group of H-bonds is formed between the side chain and the base pairs. This group couples with the H-bonds inside the base pairs. In this paper we construct a new classical model governing the dynamics of the DNA-protein recognition and the triggered breather soliton excitation that depicts the propagating local conformation. The coupling dynamics are then interpreted for various protein functions in its interaction with DNA.

\section{Model Description}
The local opening of DNA in gene recognition and regulation is triggered by nearby proteins in specific locations, while the specificity is dictated by H-bonds involved in the DNA-protein interaction. Our model should facilitate the triggering of a local conformation and its propagation through the DNA chain.

\begin{figure}
\includegraphics[width=0.41\textwidth]{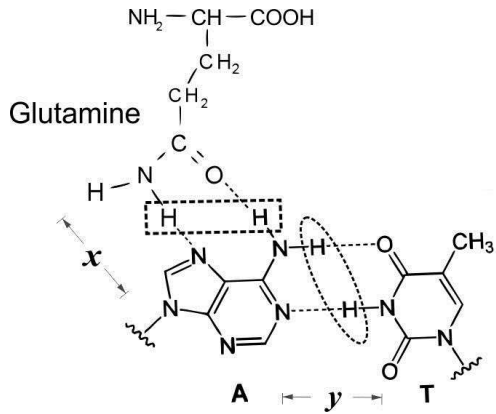}
\caption{Protein-DNA H-bonding. We pick the example case glutamine chain binding to A-T pair. The H-bonds are in box ($x$) and in oval ($y$).}
\label{illus}
\end{figure}

We assume a preexisted protein whose one of the peptide bonds is in the range of the H-bonds of its selected base pair (Fig.~\ref{illus}). By such conditions, the protein is ready to attach DNA by making additional H-bonds to the base pair. The model contains two degrees of freedom, $y_n$ and $x_m$ which correspond to the stretching of H-bonds from equilibrium in the base pairs and the stretching of H-bonds connecting protein and a DNA base respectively. The indices $n$ and $m$ indicate the base pair and protein location in DNA chain consecutively. For simplicity, we consider a sufficiently long planar DNA chain having the axis $z$ and harmonic base pairs stacking interaction. The potential for each group of H-bonds is approximated by a Morse potential. Hence, our model Hamiltonian is
\begin{equation}
H=H_\mathrm{DNA} + H_\mathrm{prot} + H_\mathrm{int},
\end{equation}
where the DNA part is
\begin{equation}
H_\mathrm{DNA}=\sum_n \frac{p_{\mathrm{y}n}^2}{2m}+D_{(\textbf{z})}(\mathrm{e}^{-\alpha y_n}-1)^2+\frac{k}{2}(y_n-y_{n-1})^2.
\end{equation}
We take homogenous mass $m$ and momentum for all base pairs and also a common coupling constant $k$ along the strand. The potential depth $D_{(z)}$ depends on the base type and its inverse width is $\alpha$, it creates the existence of a specific binding site together with $E$. The second part involves the bonds between a protein and a nearest base pair,
\begin{equation}
H_\mathrm{prot}=\frac{p_{\mathrm{x}m}^2}{2M}+E(\mathrm{e}^{-\beta x_m} - 1)^2
\end{equation}
where the potential depth $E$ and its inverse width $\beta$ depend on the H-bonds between the peptide group and the base pair. Actually, this model is ready to be applied for many proteins by specifying the multiple sites $m$. The last part describes interaction between the protein and all the bases in a rather general form, and, interaction decay can be added by specifying the decay factor $f_{mn}$,
\begin{equation}\label{int}
H_\mathrm{int}=\sum_n \frac{\chi}{2} x_m^{a}y_n^{b}f_{mn}.
\end{equation}
The coupling constant $\chi$ is a free parameter that determines the sensitivity and the strength of the interaction, the value should be set to fit the reality. It can be seen that the potential depths $D_{(z)}$ and $E$ determine the DNA-protein interaction specificity while these values are obtained numerically or by experiments such as in \cite{gao1984self}. Here we take a gaussian decay parameter,
\begin{equation}\label{decay}
f_{mn}=\exp[-\sigma^2(m-n)^2]
\end{equation}

The dynamics of the DNA-protein interaction rely heavily on the values of $a$ and $b$ in the $H_{\mathrm{int}}$. In general any integer would couple these oscillators, buts we need to consider the stability of the small amplitude dynamics of $x$ and $y$ because the model wants to provide a soliton excitation and avoid chaotic behavior in small amplitudes. To ensure this, the $H_{\mathrm{int}}$ needs to be higher than second order, $a+b>2$. One can check the implication via perturbation method if $a=b=1$ then interaction terms will exist and ruin the zeroth order equations. This will be made clear in next section. 

To give a further restriction, we consider some biological requirements, the protein should trigger a local opening in DNA even when all base pairs are still closed (i.e. $y_n=0$ initially). This requires the interaction force in the equation of motion for $y_n$, Eq.~(\ref{eomy}), does not contain the amplitude $y_n$, otherwise a closed DNA segment will remain closed even though the protein is interacting nearby. Hence we take $a=2$, and also, $b=1$ on the other hand, this will derive the desired equation of motions. 

The mathematical stability analysis of the Hamiltonian system with these settings has been conducted in \cite{donnyunpublishedstability}, the model phase portraits are given to describe the qualitative dynamics concerning the model plausibility.

\section{Nonlinear Excitations in the DNA-Protein Interaction}
\subsection{The multiscale expansion method}
To investigate how the protein induces a local base pair opening that precedes replication or regulation processes, we need to find the analytical form of the nonlinear excitations inherent to this system. A well known perturbative method has been developed in \cite{remoissenet1986low} to look for low amplitude breather solitons in nonlinear lattices. The method is actually based on the multiscale expansion technique such as in \cite{kawahara1973derivative}, the $t$ and $x$ variables are expanded to $x_0,x_1,\dots$ and $t_0,t_1,\dots$ where $x_n=\epsilon^n x$ and $t_n=\epsilon^n t$ \cite{nayfeh2004perturbation}. 

Using these multiple scales, we treat the ''fast'' and ''slow'' varying time and spatial scales separately so that we obtain envelope amplitudes and phases in different scales. A naive perturbation cannot, in general, achieve such solutions.

We first derive the equations of motion,
\begin{subequations}\label{eom}
\begin{eqnarray}
\nonumber m\ddot{y}_l & = & 2\alpha D_{(z)}(\mathrm{e}^{-2\alpha y_l}-\mathrm{e}^{-\alpha y_l})+k(y_{l+1}-2y_l\\* &&+\; y_{l-1})-\chi x_m^2f_{ml}/2, \label{eomy} \\* 
M\ddot{x}_m & = & 2\beta E(\mathrm{e}^{-2\beta x_m}-\mathrm{e}^{-\beta x_m})-\chi x_m\sum_l y_lf_{ml}. \label{eomx}
\end{eqnarray}
\end{subequations}
According to the PB approach \cite{peyrard2004nonlinear} it is assumed that the oscillations of bases are large enough to be anharmonic, but still insufficient to break the H-bond since the Morse plateau is not yet reached. The shifts $Y\equiv\alpha y$ and $X\equiv \beta x$ oscillate around the bottom of symmetric potential, hence the transformations $Y_n=\epsilon\phi_n$ and $X_m=\epsilon\psi_m$ can be safely implemented. 

To get the small amplitude solutions we expand Eqs.~(\ref{eom}) until $O(\epsilon^2)$ in terms of $\phi$ and $\psi$. Here we use the continuum approximation, assuming a long DNA chain with lattice space $a\rightarrow 0$, implying $na\rightarrow z$, $m/a\rightarrow\rho$, $ka\rightarrow K$, $D_{(z)}/a\rightarrow\mathcal{D}$, $f_{ml}\rightarrow f(z)=\exp[-\sigma^2(z-z_0)^2]$, and $\chi/a\rightarrow\mathcal{X}$ so that we get
\begin{subequations}\label{eomperturb}
\begin{eqnarray}
\nonumber \phi_{tt}- S\phi_{zz} + V_{(z)}\phi&=& V_{(z)}\left(\frac{3}{2}\epsilon\phi^2-\frac{7}{6}\epsilon^2\phi^3 +O(\epsilon^3)\right)\\*&&-\;\frac{\mu}{2}\epsilon\psi^2f(z), \\
\nonumber\psi_{tt}+W\psi &=& W\left(\frac{3}{2}\epsilon\psi^2-\frac{7}{6}\epsilon^2\psi^3+O(\epsilon^3)\right)\\*&&-\;\eta\epsilon\psi\int\phi f(z), \mathrm{d}z,
\end{eqnarray}
\end{subequations}
where the continuum parameters are
\begin{small}
\begin{equation}\label{params}
V_{(z)}=\frac{2\alpha^2\mathcal{D}}{\rho}, W=\frac{2\beta^2E}{M},  S=\frac{K}{\rho}, \mu=\frac{\mathcal{X}\alpha}{\rho\beta^2}, \eta=\frac{\mathcal{X}}{M\alpha}.
\end{equation}
\end{small}
From Eqs.~(\ref{eomperturb}) we look for perturbative solution $\phi=\phi^{(0)}+\epsilon\phi^{(1)}+\dots$ and $\psi=\psi^{(0)}+\epsilon\psi^{(1)}+\dots$. We examine solution up to order $\epsilon$ where the second harmonics appear. By the multiscale expansion, we expand the derivatives
\begin{equation}\label{multiscale}
\frac{\partial}{\partial z}=\frac{\partial}{\partial z_0}+\epsilon\frac{\partial}{\partial z_1}+\dots, \quad \frac{\partial}{\partial t}=\frac{\partial}{\partial t_0}+\epsilon\frac{\partial}{\partial t_1}+\dots.
\end{equation}
For simplicity hereafter, we write $z\equiv z_0$, $Z\equiv z_1$, $t\equiv t_0$, $T\equiv t_1$, and $\tau\equiv t_2$ and take the decay factor $f$ only dependent of $z$. The multiscale expansion gives the time or spatial derivative for each order of $\epsilon^n$ so that we can solve the equations recursively. It should be borne in mind that $\phi=\phi(z,Z,t,T,\tau)$ and $\psi=\psi(t
,T,\tau)$, the latter is not imposing explicit any spatial variables because Eq.~(\ref{eomx}) does not contain a spatial derivative by the continuum approximation. Inserting the expansions for $\phi,\psi$ and Eq.~(\ref{multiscale}) into Eqs.~(\ref{eomperturb}), we get
\small
\begin{widetext}
\begin{subequations}\label{orders}
\begin{eqnarray}
O(\epsilon^0):\;\phi^{(0)}_{tt}-S\phi^{(0)}_{zz}+V_{(z)}\phi^{(0)} &=& 0, \label{o_0e1} \\
\psi^{(0)}_{tt}+W\psi^{(0)} &=& 0, \label{o_0e2}\\
O(\epsilon^1):\;\phi^{(1)}_{tt}-S\phi^{(1)}_{zz}+V_{(z)}\phi^{(1)} &=& -2\left(\phi^{(0)}_{tT}-S\phi^{(0)}_{zZ}\right)+\frac{3}{2}V_{(z)}\phi^{(0)2}-\frac{\mu}{2}\psi^{(0)2}f, \label{o_1e1}\\ 
\psi^{(1)}_{tt}+W\psi^{(1)} &=& -2\psi^{(0)}_{tT}+\frac{3}{2}W(\psi^{(0)})^2-\eta\int\psi^{(0)}\phi^{(0)}f\,\mathrm{d}z\mathrm{d}Z, \label{o_1e2}\\
O(\epsilon^2):\;\phi^{(2)}_{tt}-S\phi^{(2)}_{zz}+V_{(z)}\phi^{(2)} &=& -2\left(\phi^{(1)}_{tT}-S\phi^{(1)}_{zZ}\right)-\left(\phi^{(0)}_{TT}-S\phi^{(0)}_{ZZ}\right)-2\phi^{(0)}_{t\tau}+3V\phi^{(0)}\phi^{(1)}-\frac{7}{6}V\phi^{(0)3}-\mu\psi^{(0)}\psi^{(1)}f, \label{o_2e1}\\
\psi^{(2)}_{tt}+W\psi^{(2)} &=& -2\psi^{(1)}_{tT}-\psi^{(0)}_{TT}-2\psi^{(0)}_{t\tau}+3W\psi^{(0)}\psi^{(1)}-\frac{7}{6}W\psi^{(0)3}-\eta\int\left(\psi^{(1)}\phi^{(0)}+\psi^{(0)}\phi^{(1)}\right)f\,\mathrm{d}z\mathrm{d}Z. \label{o_2e2}
\end{eqnarray}
\end{subequations}
\end{widetext}
\normalsize
We have now broken down the nonlinear problems into some linear homogenous equations. The solutions are
\begin{subequations}\label{sols}
\begin{eqnarray}
\phi^{(0)}&=& A_1(Z,T,\tau)\mathrm{e}^{i\theta}+\mathrm{c.c.},\\
\psi^{(0)}&=& 2(\tau)\mathrm{e}^{i\varphi}+\mathrm{c.c.},\\
\nonumber \phi^{(1)}&=& 3|A_1|^2-\frac{\mu f(z)}{2\sigma V_{(z)}}|A_2|^2-\frac{1}{2}A_1^2\mathrm{e}^{2i\theta} \\* &&+\;\frac{\mu f(z)}{12\sigma V_{(z)}}A_2^2\mathrm{e}^{2i\varphi}+\mathrm{c.c.},\\
\nonumber \psi^{(1)}&=& 3|A_2|^2-\frac{1}{2}A_2^2\mathrm{e}^{2i\varphi}+
\frac{\eta\sqrt{\pi}}{\sigma}\left[\frac{A_2\int A_1 \mathrm{d}Z}{\omega^2+2\omega\sqrt{W}}\mathrm{e}^{i(\tilde{\theta}+\varphi)}\right. \\* && \left. +\;\frac{A_2^*\int A_1 \mathrm{d}Z}{\omega^2-2\omega\sqrt{W}}\mathrm{e}^{i(\tilde{\theta}-\varphi)}\right]\mathrm{e}^{-q^2/4\sigma^2}+\mathrm{c.c.},\label{sol_d}
\end{eqnarray}
\end{subequations}
where $\theta=qz-\omega t$, $\tilde{\theta}=qz_0-\omega t$, and the phase $\varphi=\sqrt{W}t$. From Eq.~(\ref{o_0e1}) we get the dispersion relation,
\begin{equation}
\omega^ 2 = V_{(z)} + Sq^2.
\end{equation}
Next, from Eq.~(\ref{o_1e2}) we find $\partial A_2/\partial T = 0$ so $A_2$ has no dependence of $T$. Finally, the slow varying envelopes $A_1$ and $A_2$ is determined by solving the coupled NLS-like equations obtained from zeroing the secular terms ($\exp(\pm i\theta)$ and $\exp(\pm i\varphi))$ in $O(\epsilon^2)$,
\begin{subequations}\label{NLS}
\begin{eqnarray}
i\frac{\partial A_1}{\partial\tau}&+& P_1\frac{\partial^2 A_1}{\partial \xi^2}+Q_1|A_1|^2A_1= 3\mu f|A_2|^2A_1, \label{NLS1} \\*
i\frac{\partial A_2}{\partial\tau}&+& Q_2|A_1|^2A_2=\eta\gamma\int|A_1|^2 A_2\,\mathrm{d}Z, \label{NLS2}
\end{eqnarray}\end{subequations}
where
\begin{eqnarray}
\nonumber \gamma&=&\left[3+\frac{2\eta\sqrt{\pi}}{\omega^2-4W}\mathrm{e}^{-q^2/4\sigma^2}\right]\frac{\sqrt{\pi}}{\sigma},\quad Q_1=4V_{(z)}, \\
Q_2&=& \left[4W+\frac{5\mu\eta\sqrt{\pi/2}}{12\sigma^2 V_{(z)}}\right],\quad P_1=\frac{S-V_g^2}{2\omega},
\end{eqnarray}
and $\xi=Z-V_gT$ is a right-moving coordinate having group velocity $V_g=Sq/\omega$. The integral of $|A_1|^2$ with respect to $Z$ over entire space is a finite function of time because we assume a localized solitonic wave.

\subsection{Nonlinear Excitations}
The bright soliton solutions of coupled NLS have been fairly investigated, such as in \cite{radhakrishnan1995bright}. We use the Hirota bilinear method \cite{hirota2004direct} to solve Eqs.~(\ref{NLS}) with the transformations
\begin{equation}\label{hirotatrans}
A_1\equiv \frac{G(\xi,\tau)}{F(\xi,\tau)} \quad \mathrm{and} \quad A_2\equiv \frac{H(\xi,\tau)}{F(\xi,\tau)}\bigg|_{\xi=\xi_0}
\end{equation}
where $F\in\mathbb{R}$ and $G,H\in\mathbb{C}$. Here we apply a technique that assumes spatial dependency of $A_2$ in the first place and discarding it by inserting a constant $\xi=\xi_0$ after the solution is found. By inserting Eq.~(\ref{hirotatrans}) into Eqs.~(\ref{NLS}) we get the bilinear forms
\begin{subequations}\label{bilinear}
\begin{eqnarray}
\nonumber \left(iD_\tau+P_1D^2_\xi\right) G \cdot F &=& 0, \\* 
Q_1|G|^2 - \mu\gamma f |H|^2 &=& P_1 D_\xi^2 F \cdot F, \label{bilinear1} \\*
\nonumber iD_\tau H \cdot F &=& 0, \\*
\qquad Q_2|H|^2 &=& \eta\gamma|G|^2. \label{bilinear2}
\end{eqnarray}
\end{subequations}
From Eq.~(\ref{bilinear2}) we can relate $A_1$ and $A_2$ by $|H|^2=\eta\gamma|G|^2/Q_2$. The problem of finding one-soliton solution with a common phase is now equivalent with solving one NLS in the form of
\begin{equation}
i\frac{\partial A_1}{\partial\tau}+ P_1\frac{\partial^2 A_1}{\partial \xi^2}+Q'(z)|A_1|^2A_1=0,
\end{equation}
where
\begin{equation}
Q'(z)=Q_1-\frac{3\mu\eta\gamma}{Q_2}f(z)
\end{equation}
is obtained from Eq.~(\ref{bilinear2}). The solution for $PQ'>0$, the bright soliton case, is \cite{remoissenet1986low}
\begin{eqnarray}\label{A_1}
\nonumber A_1(\xi,\tau)&=& A\, \mathrm{sech}\left[A\,\left(\frac{Q'}{2P_1}\right)^{1/2} \left(\xi-\frac{v_e\tau}{P_1}\right) \right] \\*
&& \times \exp\left[i\left(\frac{v_e}{2P_1}\right) \left(\xi-\frac{v_c\tau}{P_1}\right) \right]
\end{eqnarray}
where $A$ is the amplitude,
\begin{equation}\label{ampNLS}
A(z)=\left(\frac{v_e^2 - 2v_ev_c}{2P_1Q'(z)}\right)^{1/2},
\end{equation}
with conditon $v_e^2 - 2v_e v_c >0$, and $v_e,v_c$ are the envelope and carrier waves velocity while we take $v_c=gv_e$ for positive $g$. By taking a common phase, we get that the expression for $A_2$ just differs by the amplitude according to Eq.~(\ref{bilinear2}),
\begin{equation}\label{A_2}
A_2(\tau)=\left(\frac{\eta\gamma}{Q_2}\right)^{1/2} A_1(\xi,\tau)|_{\xi=\xi_0,z=z_0}.
\end{equation}

To obtain the solution, we first calculate the integral term in Eq.~(\ref{sol_d}) by changing the domain $Z$ to $\xi$,
\begin{equation}
\int_{-\infty}^{\infty} A_1\,\mathrm{d}Z = CA\exp\left[\frac{iv_e^2 }{2P_1^2}(1-g)\tau\right]
\end{equation}
where
\begin{equation}
C=\frac{2\pi P_1}{v_e\sqrt{1-2g}}\,\mathrm{sech}\left(\frac{\pi}{2\sqrt{1-2g}}\right).
\end{equation}
Inserting Eqs.~(\ref{A_1}) and (\ref{A_2}) into Eqs.~(\ref{sols}) and setting $\xi_0$ and $z_0$ to zero, we get
\begin{subequations}\label{sol_full}
\begin{eqnarray}
\nonumber Y(z,t)&=& \epsilon 2A \,\mathrm{sech}\,\Theta \cos(N_+ t)+ \epsilon^2 A^2\, \mathrm{sech}^2\Theta \\* && \nonumber\times\left\{3-\cos 2(Qz-Mt)-\Lambda(z)\right.\\* &&\left.\times\; [1-\cos2((Q-q)z-Mt)] \right\} +O(\epsilon^3), \label{sol_full_1} \\*
\nonumber X(t) &=& \epsilon 2A \left(\frac{\eta\gamma}{Q_2}\right)^{1/2} \mathrm{sech}\, \Theta_0 \cos(N_+ t) + \epsilon^2 A^2 \left(\frac{\eta\gamma}{Q_2}\right) \\* && \nonumber \times\;\mathrm{sech}^2 \Theta_0 \,[3 - \cos 2(N_+ t)] + \epsilon^2\frac{\eta\sqrt{\pi}}{\sigma} \mathrm{e}^{-q^2/4\sigma^2} \\* && \nonumber \times\; 2CA\left(\frac{\eta\gamma}{Q_2}\right)^{1/2} \mathrm{sech}\, \Theta_0[\omega_+^{-1}\cos(N_+ +\omega)t \\* &&  +\;\omega_-^{-1}\cos(N_- +\omega)t]+O(\epsilon^3), \label{sol_full_2}
\end{eqnarray}
\end{subequations}
where
\begin{eqnarray}
\Theta(z,t) &=& \epsilon\frac{v_e\sqrt{1-2g}}{2P_1}\left(qz-(V_g+\epsilon\frac{v_e}{P_1})t\right), \\*
\Theta_0(t) &=& \epsilon^2 \frac{v_e\sqrt{1-2g}}{2P_1}t, \\*
Q &=& q +\epsilon \frac{v_e}{2P_1}, \\*
M &=& \omega+\epsilon\frac{v_e}{2P_1}\left[V_g+\epsilon\left(\frac{g v_e}{P_1}\right)\right], \\*
N_\pm&=&\pm\sqrt{W} +\epsilon^2 g \frac{v_e^2}{2P_1^2}, \\*
\Lambda(z)&=& \frac{\mu\eta\gamma}{V_{(z)}Q_2}\mathrm{e}^{-\sigma^2 z^2}, \\* 
\omega_\pm&=&\omega^2 \pm 2\omega\sqrt{W}.
\end{eqnarray}

One can see from Eq.~(\ref{sol_full}), if the coupling $\chi=0$ then $\Lambda(z)=0$, hence $y(z,t)$ will be identical with the Peyrard-Bishop breather solution \cite{peyrard2004nonlinear} while $x(t)=0$ as if there is no interacting protein. 
\begin{figure}[t]
\includegraphics[width=0.5\textwidth]{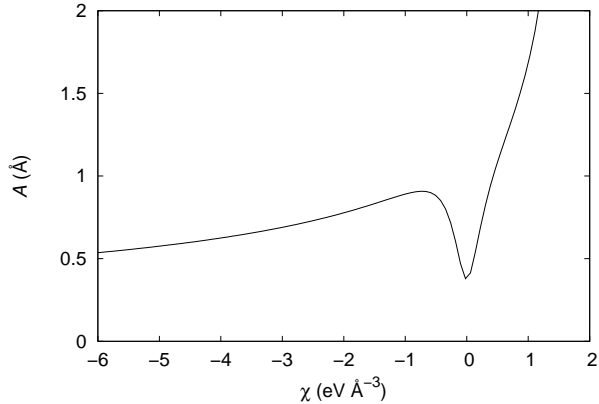}
\caption{Relation between amplitude $A$ and the coupling constant}
\label{A_vs_chi}
\end{figure}

We take and adjust the PB parameters from \cite{zdravkovic2009parameter},
\begin{eqnarray}\label{parameters}
\nonumber \alpha &=& 1.2 \sqrt{2} \,\mathrm{\AA^{-1}},\quad D=0.07 \,\mathrm{eV},\quad k=12\,\mathrm{N/m}, \\*
q&=& 0.18 \,\mathrm{\AA^{-1}},\quad g=0.47,\quad v_e=1888 \,\mathrm{m/s},
\end{eqnarray}
here we take constant $D$ for simplicity. The length between two base pairs is $a=3.4\,\mathrm{\AA}$ and the nucleotide mass is $m=5.1\times 10^{-25}\,\mathrm{kg}$. Here the value of $q$ corresponds to a wavelength covering 10 basepairs. We take $E=D$ and $\beta=\alpha$ as the connecting hydrogen bonds between glutamine and adenine are the same as those connecting A-T. We interpret the decay factor $\sigma/\sqrt{2}$ as the inverse width of the protein, an $\alpha$-helix protein is $12\,\mathrm{\AA}$ in diameter \cite{matsushima1997three} hence $\sigma=0.117\,\mathrm{\AA}^{-1}$. The protein effective mass $M$ is rather free because any proteins could have an arbitrary amount of amino acid sequences. However in this case we take it as a glutamic acid weight, $M=2.47\times 10^{-24}\,\mathrm{kg}$. We are left with a free parameter which is the coupling constant $\chi$ whose value should have significance in the dynamics of the interaction.

\section{Discussions and Conclusions}
To discuss the model relevancy with biological reality, we examine the restrictions of our introduced coupling constant $\chi$ and its ramification with the amplitudes. The value of the coupling is central to the complex interaction and should have various interpretations. To see this consider Eq.~(\ref{ampNLS}), taking $z=z_0$, we get the restriction
\begin{equation}
Q_1Q_2 - 3\mu\eta\gamma > 0,
\end{equation}
or more explicitly, recall that $\mathcal{X}=\chi/a$,
\begin{equation}
Q_1C + (Q_1D - A)\mathcal{X}^2 - B\mathcal{X}^3 > 0
\end{equation}
where
\begin{eqnarray}
A &=& \frac{9\sqrt{\pi}}{M\rho\beta^2\sigma},\quad B = \frac{6\sqrt{\pi}}{M\alpha(\omega^2-4W)}\mathrm{e}^{-q^2/4\sigma^2}, \\*
C &=& 4W, \quad\qquad D = \frac{15\sqrt{\pi/2}}{12M\rho\beta^2\sigma^2V_{(z)}}.
\end{eqnarray}
If $Q_1D - A >0$ then we require
\begin{equation}
\sigma < 0.392\,\mathrm{\AA^{-1}},
\end{equation}
meaning the protein diameter $\sqrt{2}/\sigma > 3.59\,\mathrm{\AA}$ which is certainly the case because it is barely a base pair length. Now we are left with $B$, whose sign is determined by $\omega^2-4W$. The positive and negative case are respectively
\begin{equation}
M>8\beta^2E/\omega^2, \quad M<8\beta^2E/\omega^2.
\end{equation}
A positive $B$ will produce a relation between the amplitude $A$ and coupling $\chi$ as in Fig.~\ref{A_vs_chi}, the main property is that $A\rightarrow0$ for highly negative $\chi$. In contrast, a negative $B$ will vertically mirror the relation i.e. $A\rightarrow0$ for highly positive $\chi$. Here our choice falls in the positive case. We will discuss the four interesting values in Fig.~\ref{A_vs_chi} of $\chi$: zero, local maximum, approaching singularity, and highly negative.

\begin{figure*}
\centering
\includegraphics[width=0.78\textwidth]{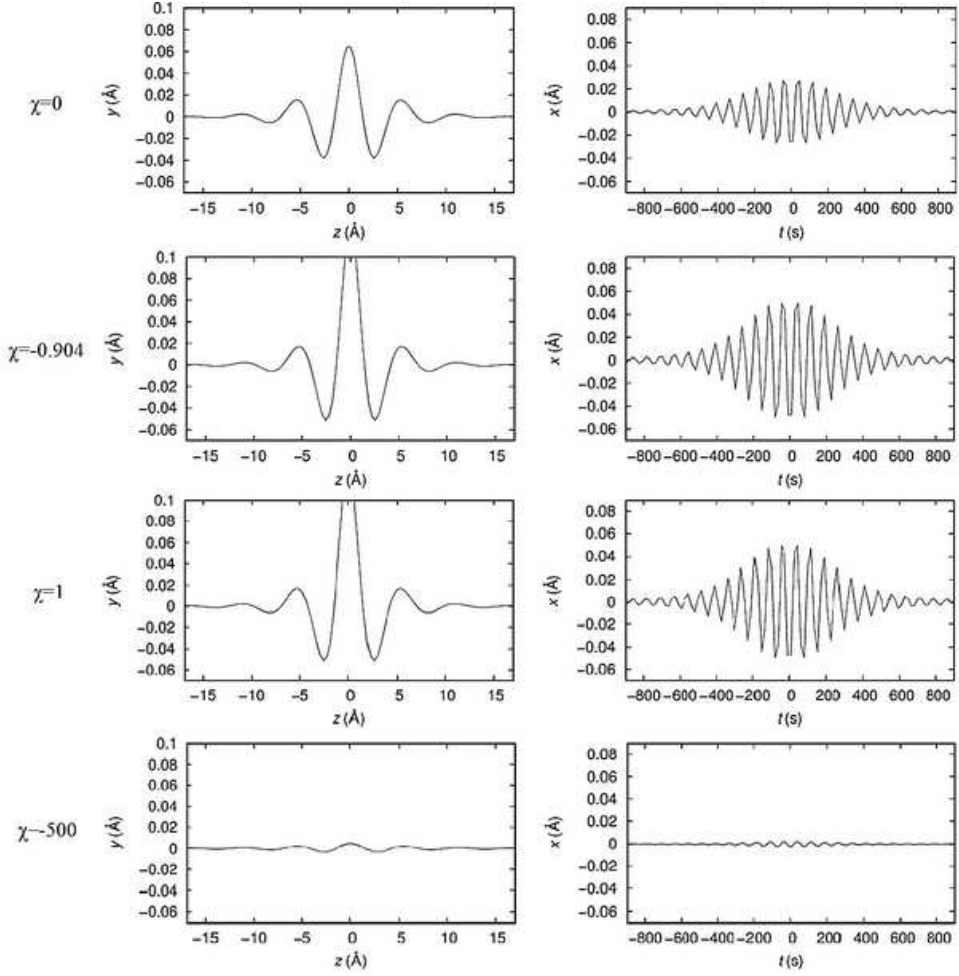}
\caption{Plot of the solutions with coupling constant $\chi=0$, $\chi=-0.904$ (local maximum), $\chi=1$ (approaching singularity), and $\chi=-500$ (highly negative).}
\label{amp_y}
\end{figure*}

\begin{figure*}
\includegraphics[width=\textwidth]{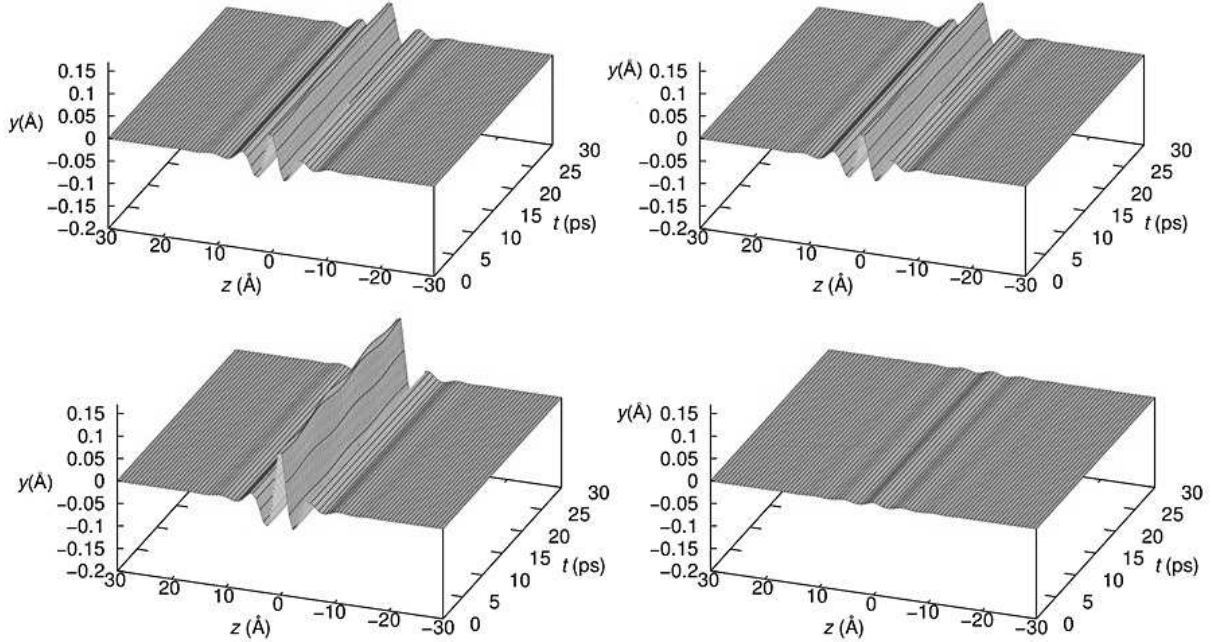}
\caption{Plot of the strecth $y$ with respect to DNA chain $z$ and time, (top left) $\chi=0$, (top right) $\chi=-0.904$, (bottom right) $\chi=1$, and (bottom left) $\chi=-500$.}
\label{amp_y_3D}
\end{figure*}

For $\chi=0$ it is shown in Eqs.~(\ref{sol_full}), Fig.~\ref{amp_y}, and Fig.~\ref{amp_y_3D} that the solution is identical to the PB breather while the H-bonds between protein and DNA are not shifted. However, if we take $\chi$ near zero, the DNA is opened while the protein side chain is undergoing a local oscillation. The interaction triggers a local base pairs opening to let the protein recognize a specific sequence of base pairs while previously the bases are being hindered from outside world by the DNA backbone. Our model only contains the bond shifts and thus has a radial symmetry, this lets the bases twist out while being recognized. The moving breather soliton is interpreted as the mediator of the allosteric transmissions in DNA \cite{hogan1979transmission,ptashne1985gene} that facilitate the long-range information transfer between two vastly separated specific DNA-binding proteins. 

Our result for small $\chi>0$ is in agreement with the statistical model in \cite{sataric2002impact} that implies a breather excitation or amplification of an already existing breather. We should restrict $\chi$ in the value where the amplitudes vary linearly with $\chi$. In addition, from Eq.~(\ref{sol_full_1}) we find that the wavelength near the protein is reduced by $q$ in the $\Lambda(z)$ term. We predict from Eq.~(\ref{sol_full_2}) that the protein will sustain a small local oscillation that reduces over time, by the slowly varying envelope wave over 1000 ps length. These type of vibration is mechanically due to the recoil from the base opening process.

The singular point is due to the square root in Eq.~(\ref{ampNLS}). We cannot interpret this as a totally denatured or separated DNA strands because the case is outside our small amplitude approximation. On the other hand, we have not found the significance of the local maximum case of $\chi$ other than its relatively high amplitude property. The local maximum occurs when $Q_1D-A>0$. For $Q_1D\approx A$ or for very small proteins, the local maximum is ceased.

The case of highly negative $\chi$ is particularly interesting because, contrary to the previous cases, it totally reduces the amplitudes. Here the protein can practically close the base pairs within its reach, which is restricted because of $f(z)$. This contributes to the DNA recombination mechanism or the reverse of denaturation. The naturally occurring base pairs closing is the renaturation process catalysed by proteins such as RAD1O~\cite{sung1992renaturation}, while a similar mechanism can be engineered by DNA hybridization. The base recombination process is central to the polymerase chain reaction which is widely used for DNA testing.

At last, the value of effective mass $M$ is still unclear because it depends on the geometrical features of the protein itself and the interaction it conducts. A further investigation is needed because the overall interaction dynamics can be extremely different as the coupling behavior is dependent on $M$. It is also interesting to study the base pair zippering by protein because it needs to regularize the thermal fluctuation that forces the bases to open again after being closed by the renaturation or hybridization processes.

\begin{acknowledgments}
The authors would like to thank Ministry of Research Technology and Higher Education of Republic of Indonesia for research funding Desentralisasi 2016. DD thanks the members of Theoretical Physics Laboratory ITB for hospitality.
\end{acknowledgments}


\bibliography{apssamp}

\end{document}